
\input harvmac
\def\L{{\cal L}}
\def\O{{\cal O}}
\def\l{\lambda}

\def\bu{\bar u}
\def\bd{\bar d}
\def\bell{\bar\ell}
\def\gsim{\ \rlap{\raise 2pt \hbox{$>$}}{\lower 2pt \hbox{$\sim$}}\ }
\def\lsim{\ \rlap{\raise 2pt \hbox{$<$}}{\lower 2pt \hbox{$\sim$}}\ }

\Title{hep-ph/9504312, WIS-95/18/Apr-PH}
{\vbox{\centerline{Gauge Unification, Yukawa Hierarchy and the
$\mu$ Problem}}}
\bigskip
\centerline{Yosef Nir}
\smallskip
\centerline{\it Department of Particle Physics}
\centerline{\it Weizmann Institute of Science, Rehovot 76100, Israel}
\bigskip
\baselineskip 18pt

\noindent
The hierarchy in the Yukawa couplings may be the result of a gauged
horizontal $U(1)_H$ symmetry. If the mixed anomalies of the Standard
Model gauge group with $U(1)_H$ are cancelled by a Green-Schwarz
mechanism, a relation between the gauge couplings, the Yukawa couplings
and the $\mu$-term arises. Assuming that at a high energy scale
$g_3^2=g_2^2={5\over3}g_1^2$ and $(m_e m_\mu m_\tau)/(m_d m_s m_b)\sim
\lambda$ (where $\lambda$ is of the order of the Cabibbo angle),
the $U(1)_H$ symmetry solves the $\mu$-problem with
$\mu\sim\lambda m_{3/2}$.

\Date{4/95}

A possible intriguing relation between hierarchies in the fermion
Yukawa couplings and unification of gauge couplings has been
recently proposed by Bin\'etruy and Ramond [BR]
\ref\BiRa{P. Bin\'etruy and P. Ramond, LPTHE-ORSAY 94/115,
hep-ph/9412385.}\
and by Ib\'a\~nez and Ross
\ref\IbRo{L. Ib\'a\~nez and G.G. Ross, Phys. Lett. B332 (1994) 100.}.
The relation arises if anomalies of a gauged horizontal
$U(1)_H$ symmetry are cancelled by a Green-Schwarz mechanism
\ref\Iban{L.E. Ib\'a\~nez, Phys. Lett. B303 (1993) 55.}.
One of the basic assumptions of BR was that the $\mu$-term in
the superpotential is neutral under $U(1)_H$. In this work, we
investigate the more general case, namely that the $\mu$-term
carries an arbitrary charge. We point out that if we take that,
at a high scale, (a) the gauge couplings satisfy
\eqn\unif{g_3^2=g_2^2={5\over3}g_1^2,}
and (b) the fermion masses satisfy
\eqn\hier{{m_em_\mu m_\tau\over m_dm_sm_b}\sim\lambda}
(where $\l\sim0.2$ is the small breaking parameter of $U(1)_H$), then
the horizontal symmetry solves the $\mu$-problem with
\eqn\solvemu{\mu\sim \l m_{3/2}.}

We work in the framework of Supersymmetry and
$SU(3)_C\times SU(2)_L\times U(1)_Y\times U(1)_H$ gauge symmetry.
We now list the assumptions that lead to our results.
(We also show that some of the assumptions made by BR are actually
not necessary for their result.)

1. {\it The smallness and hierarchy among the Yukawa
couplings is a result of a horizontal $U(1)_H$ symmetry that is
broken by a single small parameter $\l(\sim0.2)$ whose charge
under $U(1)_H$ is defined to be --1}.\foot{
The parameter $\l$ is the ratio between the VEV of a SM-singlet
scalar field that carries charge --1 under $U(1)_H$ and a somewhat
higher scale where the information about $U(1)_H$ breaking is
communicated to the light fermions
\ref\FrNi{C.D. Froggatt and H.B. Nielsen, Nucl. Phys. B147 (1979) 277.}.}

It follows that the order of magnitude of the various Yukawa couplings,
\eqn\Yukawa{\L^{\rm Y}=\l^u_{ij}Q_i\bar u_j\phi_u+
\l^d_{ij}Q_i\bar d_j\phi_d+\l^\ell_{ij}L_i\bar\ell_j\phi_d,}
can be estimated by the following selection rules
\ref\lnsa{M. Leurer, Y. Nir and N. Seiberg,
 Nucl. Phys. B398 (1993) 319.}:
\eqn\selection{\eqalign{
\l^u_{ij}\sim&\ \cases{\l^{H(Q_i)+H(\bu_j)+H(\phi_u)}
&$H(Q_i)+H(\bu_j)+H(\phi_u)\geq0$,\cr
0&$H(Q_i)+H(\bu_j)+H(\phi_u)<0$,\cr}\cr
\l^d_{ij}\sim&\ \cases{\l^{H(Q_i)+H(\bd_j)+H(\phi_d)}
&$H(Q_i)+H(\bd_j)+H(\phi_d)\geq0$,\cr
0&$H(Q_i)+H(\bd_j)+H(\phi_d)<0$,\cr}\cr
\l^\ell_{ij}\sim&\ \cases{\l^{H(L_i)+H(\bell_j)+H(\phi_d)}
&$H(L_i)+H(\bell_j)+H(\phi_d)\geq0$,\cr
0&$H(L_i)+H(\bell_j)+H(\phi_d)<0$.\cr}\cr}}
The zeros, corresponding to negative charges, are a result of the
holomorphy of the superpotential \lnsa.
If there are no zero eigenvalues (as experimentally known for quarks and
leptons, with the possible exception of $m_u=0$
\ref\BSN{For a recent review, see T. Banks, Y. Nir and N. Seiberg,
hep-ph/9403203, to appear in the Proc. of the Yukawa Couplings and
the Origins of Mass Workshop.}), we get
\eqn\detMf{\eqalign{
\det M^u\sim&\ \vev{\phi_u}^3\l^{\sum_i[H(Q_i)+H(\bu_i)]+3H(\phi_u)},\cr
\det M^d\sim&\ \vev{\phi_d}^3\l^{\sum_i[H(Q_i)+H(\bd_i)]+3H(\phi_d)},\cr
\det M^\ell\sim&\ \vev{\phi_d}^3\
\l^{\sum_i[H(L_i)+H(\bell_i)]+3H(\phi_d)}.\cr}}
Note that \detMf\ requires neither that third generation
fermions acquire masses without horizontal suppression,
nor that all excess charges in $\l^f_{ij}$ are positive
(two assumptions made by BR).
It only requires that all fermion masses are non-zero, and so
it holds in any phenomenologically acceptable model.

2. {\it The only fields that are in chiral representations of $U(1)_H$
and transform non-trivially under the SM gauge group are the MSSM
supermultiplets}.\foot{
The fields required by the Froggatt-Nielsen mechanism \FrNi\
are either SM-singlets or in vector representations, consistent with
this assumption.}

This allows one to calculate
the mixed anomalies $C_n$ in $SU(n)^2\times U(1)_H$,
(in $n=3,2,1$ we refer to the $SU(3)_C\times SU(2)_L\times U(1)_Y$
factors in the SM gauge group) in terms of the $H$-charges carried by
the MSSM fields only. Explicitly \BiRa:
\eqn\anomalies{\eqalign{
C_3=&\ \sum_i[2H(Q_i)+H(\bu_i)+H(\bd_i)],\cr
C_2=&\ \sum_i[3H(Q_i)+H(L_i)]+H(\phi_u)+H(\phi_d),\cr
C_1=&\ \sum_i[{1\over3}H(Q_i)+{8\over3}H(\bu_i)+{2\over3}H(\bd_i)
+H(L_i)+2H(\bell_i)]+H(\phi_u)+H(\phi_d).\cr}}

Eqs. \detMf\ and \anomalies\ lead to two independent relations between
the anomalies $C_i$, the determinants of the fermion mass matrices,
the VEVs of the Higgs doublets and the sum of Higgs charges
\eqn\defHphi{H(\phi)\equiv H(\phi_d)+H(\phi_u).}
These are
\eqn\relationa{
(\det M^u)(\det M^d)\sim\vev{\phi_u}^3\vev{\phi_d}^3\l^{C_3+3H(\phi)},}
\eqn\relationb{
(\det M^\ell)(\det M^u)^{4\over3}(\det M^d)^{1\over3}\sim
\vev{\phi_u}^4\vev{\phi_d}^4\l^{{1\over2}(C_1+C_2)+3H(\phi)}.}
The reason that there are only two independent relations is that the
Yukawa couplings have an accidental $U(1)_B\times U(1)_L$ symmetry
($B$ and $L$ stand here for Baryon and Lepton number, respectively);
only $C_3$ and $C_1+C_2$ are $U(1)_B\times U(1)_L$ invariant.

The Yukawa sector has yet another accidental symmetry, $U(1)_X$,
with $X(\phi_d)=-X(\bd_i)=-X(\bell_i)$ and all other supermultiplets
carrying $X=0$. One combination of the anomalies is $U(1)_X$ invariant
and should, therefore, appear in an $H(\phi)$-independent relation.
Indeed, dividing \relationb\ by \relationa, we get
\eqn\indepH{{(\det M^\ell)(\det M^u)^{1/3}\over(\det M^d)^{2/3}}
\sim\vev{\phi_u}\vev{\phi_d}\l^{(C_1+C_2-2C_3)/2}.}
The LHS of this relation can be estimated to be $\lsim\O(m_s m_c)$
(assuming approximate geometrical hierarchies and $\det M^\ell\lsim
\det M^d$), while the RHS is (for $\tan\beta\lsim m_t/m_b$)
$\gsim\O(m_b m_t)\l^{(C_1+C_2-2C_3)/2}$. The conclusion
is then that $C_1+C_2-2C_3>10$ (for $\l\sim0.2$): the mixed anomalies
cannot vanish simultaneously. This conclusion is independent
of $H(\phi)$. If $U(1)_H$ is a local symmetry, there should exist a
mechanism to cancel these anomalies.

3. {\it The mixed anomalies are cancelled by a Green-Schwarz mechanism}
\ref\GrSc{M. Green and J. Schwarz, Phys. Lett. B149 (1984) 117.},
{\it with Im$(S)$ playing the role of the $U(1)_H$ axion ($S$ stands
for the dilaton supermultiplet)}.

This is possible only if \Iban\
\eqn\sucessGS{{C_1\over k_1}={C_2\over k_2}={C_3\over k_3}=
\delta_{GS},}
where $k_i$ are the Kac-Moody levels and $\delta_{GS}$ is a constant
that gives the transformation law of the axion under $U(1)_H$.
In general, $k_2$ and $k_3$ are integers and $k_1$ is a rational number.
The gauge couplings for the SM gauge group are given (at the string
scale) by
\eqn\gsubi{g_i^2={g^2\over k_i}.}
We can then relate the mixed anomalies to the gauge couplings:
\eqn\anocou{C_i={g^2\delta_{GS}\over g_i^2}.}

We next turn our attention to another relation that results from
\relationa\ and \relationb\ and
is independent of $\vev{\phi_u}$, $\vev{\phi_d}$:
\eqn\indepp{{\det M^d\over\det M^\ell}\sim\l^{-{1\over2}[C_1+C_2-
{8\over3}C_3]+H(\phi)}.}
Using \anocou\ we can rewrite \indepp\ as
\eqn\indeppp{{\det M^d\over\det M^\ell}\sim\l^{-{g^2\delta_{GS}\over2}
\left[{1\over g_1^2}+{1\over g_2^2}-{8\over3g_3^2}\right]+H(\phi)}.}

Before proceeding with the analysis of \indeppp,
we would like to discuss the relation between
$H(\phi)$ and the $\mu$-problem. $H(\phi)$ is the horizontal charge
carried by the $\mu$ term in the superpotential,
\eqn\mudef{\mu\phi_u\phi_d.}
The $\mu$-term is special in that it can come from two sources:
it may arise from the high energy superpotential, in which case it is
holomorphic and its natural scale is unknown
(but most likely $\sim M_{Pl}$); or it may arise from the
K\"ahler potential, in which case it is not holomorphic and its
natural scale is the SUSY breaking scale
\ref\GiMa{G.F. Giudice and A. Masiero, Phys. Lett. B206 (1988) 480;\hfill
\break V.S. Kaplunovsky and J. Louis, Phys. Lett. B306 (1993) 269.}.
In other words, the selection rule for $\mu$ is
\eqn\selectmu{\mu\sim\cases{
\tilde\mu\l^{H(\phi)}&$H(\phi)\geq0$,\cr
m_{3/2}\l^{|H(\phi)|}&$H(\phi)<0$,\cr}}
where $\tilde\mu$ is the unknown natural scale of $\mu$
(most likely $\tilde\mu\sim M_{Pl}$).
This suggests that the
horizontal symmetry may actually solve the $\mu$-problem: if the
charge carried by the $\mu$-term is negative, then it cannot come
{}from the superpotential and its natural scale is $m_{3/2}$. Note,
however, that the charge cannot be too negative, because $\mu\ll m_{3/2}$
is phenomenologically unacceptable. With $\lambda\sim0.2$,
we can allow $H(\phi)=-1$, which would give
$\mu/m_{3/2}\sim\l$. We conclude that a continuous horizontal symmetry
can solve the $\mu$-problem. The solution requires an almost unique
choice of charge, namely that the $\mu$-term carries one negative unit of
the horizontal charge.

Eq. \selectmu\ gives a new interpretation of \indeppp;
it is a relation between Yukawa couplings, the
gauge couplings and the $\mu$-term. For example, if $H(\phi)<0$,
\indeppp\ can be rewritten as
\eqn\indepppp{{\mu\over m_{3/2}}
{\det M^d\over\det M^\ell}\sim\l^{-{g^2\delta_{GS}\over2}
\left[{1\over g_1^2}+{1\over g_2^2}-{8\over3g_3^2}\right]}.}

BR made the following further three assumptions:
\item{(i)} $H(\phi)=0$.
\item{(ii)} $k_2=k_3$, namely $g_2=g_3$.
\item{(iii)} $m_e m_\mu m_\tau\sim m_d m_s m_b$ (at the high scale).

Then \indeppp\ leads to the interesting result
$\sin^2\theta_W\equiv{g_1^2\over g_1^2+g_2^2}={3\over8}$,
in perfect agreement with extrapolated phenomenology.
The weakest assumption in this derivation is that of $H(\phi)=0$
which is just the rather arbitrary ansatz that $\mu$ is unsuppressed
relative to $\tilde\mu$ (which is unknown).

The way in which we will proceed is
to use the extrapolated phenomenology to make assumptions about
the gauge couplings and the Yukawa couplings and find the consequences
for $H(\phi)$ and the $\mu$-term. Specifically, we make the
following assumptions, which are based on running the various
gauge and Yukawa couplings to the high scale:

4. {\it At the string scale, the gauge couplings satisfy}
\eqn\unif{g_3^2=g_2^2={5\over3}g_1^2.}

5. {\it At the string scale, the fermion masses satisfy}
\eqn\hierar{{m_e\over m_\mu}\sim\l^3,\ \ {m_\mu\over m_\tau}\sim\l^2,\ \
{m_d\over m_s}\sim\l^2,\ \ {m_s\over m_b}\sim\l^2.}
(Here, we differ from BR; they take ${m_e\over m_\mu}\sim\l^2$.)
Consequently, if at the high scale $m_b$ and $m_\tau$ approximately
unify, namely $m_\tau\sim m_b$, we get
\eqn\hier{{m_em_\mu m_\tau\over m_dm_sm_b}\sim\lambda.}
This leads, through eq. \indeppp, to
\eqn\requiredH{H(\phi)=-1,}
which is just the right value to solve the $\mu$-problem (see \indepppp)
\eqn\solvemu{\mu\sim \l m_{3/2}.}

To summarize: if anomalies in a gauged horizontal $U(1)$ symmetry are
cancelled by a GS mechanism, then interesting relations among
gauge couplings, Yukawa couplings and the $\mu$ term arise.
The values of the gauge and Yukawa couplings, when extrapolated
from their measured low energy values, imply that the scale of
the $\mu$ term is below (but not far below) the SUSY breaking scale.

While this paper was in writing, we received a preprint by Dudas,
Pokorski and Savoy
\ref\DPS{E. Dudas, S. Pokorski and C.A. Savoy, SPhT Saclay T95/027
(1995), hep-ph/9504292.}\ that also
investigates the general $H(\phi)$ case.

{\bf Acknowledgments}: I am grateful to Adam Schwimmer for
many useful discussions.
YN is supported in part by the United States -- Israel Binational
Science Foundation (BSF), by the Israel Commission for Basic Research
and by the Minerva Foundation.

\listrefs
\end